\documentstyle[bo99,epsfig]{article}

\title{X-ray sources as tracers of the large-scale structure in the Universe}

\author{X. Barcons $^1$, F.J. Carrera $^{2,1}$, M.T. Ceballos$^1$, 
S. Mateos $^{1,3}$}                                                       
\affil{1) Instituto de F\'\i sica de Cantabria (CSIC-UC), 39005 Santander,
Spain\\                                                
2) Mullard Space Science Laboratory, University College London, UK\\
3) Departamento de F\'\i sica Moderna, Universidad de Cantabria, 39005
Santander, Spain}                                                
\begin{document}

\maketitle

\begin{abstract}
We review the current status of studies of large-scale structure in
the X-ray Universe.  After motivating the use X-rays for cosmological
purposes, we discuss the various approaches used on different angular
scales including X-ray background multipoles, cross-correlations of the
X-ray background with galaxy catalogues, clustering of X-ray selected
sources and small-scale fluctuations and anisotropies in the X-ray
background. We discuss the implications of the above studies for the
bias parameter of X-ray sources, which is likely to be moderate for
X-ray selected AGN and the X-ray background ($b_X\sim 1-2$).  We
finally outline how all-sky X-ray maps at hard X-rays and medium
surveys with large sky coverage could provide important tests for the
cosmological models.

\end{abstract}

\section{Introduction}

In the current cosmological picture, galaxies, clusters and
large-scale structures have grown from initial small perturbations in
the density of the Universe via gravitational collapse.  Cosmological
models are required to meet two basic observational constraints: on
the one hand the Universe at $z\sim 1500$ was very smooth, as the
cosmic microwave background (CMB) is seen to have anisotropies of
amplitude $\sim 10^{-5}$; on the other hand local mass inhomogeneities
measured through the distribution of galaxies exhibit fluctuations of
the order $\sim 1$ on scales $\sim 10\, {\rm Mpc}$.  Different
cosmologies, however, predict highly discrepant ways in which
structures on different scales grow up to the current state from the
CMB initial conditions.  The largest discrepancies occur at redshifts
$z\sim 1-5$ which is when galaxies began to collapse and to form
stars. Accessing these intermediate redshifts will provide crucial
tests for the cosmological models.

The isotropy of the cosmic X-ray background (XRB) on large angular
scales (${\Delta I\over I}$ less than a few \% on scales of degrees
and larger) suggests that most of the X-ray photons we receive
from the Universe must have been originated in the distant Universe.
Surveys at different depths carried out with $ROSAT$ have revealed
that 50-70\% of the (soft) XRB is resolved into point sources, mostly
Active Galactic Nuclei (AGN) of different classes.  Although there are
still some discrepancies in the determination of the X-ray luminosity
function and its redshift evolution, there is no doubt that most of
the XRB originates at redshift $z>1$.  Boyle et al (1994) and Page et
al (1996) who find their samples of X-ray selected AGN consistent with
pure luminosity evolution models, predict a peak in the X-ray volume
emissivity around $z\sim 1.5-2$.  Miyaji et al (1998) instead find
better consistency with luminosity dependent density evolution, in
which case the X-ray volume emissivity in AGN more luminous than
$10^{44.5}\, {\rm erg}\, {\rm s}^{-1}$ (which for the broken power-law
shape of the luminosity function account for most of the X-rays
emitted by AGN) rises steeply from $z=0$ to $z=1-2$ with no evidence
for a decline at higher redshifts. In both cases it is clear that soft
X-ray emission from the extragalactic sky comes mostly from redshifts
$z=1-2$ or larger, in a situation very similar to the star formation
in the Universe (Madau et al 1996, Boyle \& Terlevich 1998). Studying the
X-ray Universe is then likely to provide a major handle to understand
the evolution of the Universe at intermediate redshifts and therefore
it is an issue of prime cosmological relevance.

There are other reasons to prefer X-rays to carry out cosmological
studies. On the one hand the high-latitude X-ray sky is `clean', at
least at photon energies above 2 keV, galactic absorption has
negligible effects and the contribution of the Galaxy to the XRB is
less than a few \% (Iwan et al 1982).  A further reason is the small
content in stars of high galactic latitude surveys, ranging from 25\%
at bright fluxes down to probably less than 10\% at the faintest
fluxes.  

In this paper we review the current status of studies of the
large-scale structure of the Universe, which up to now has produced
relevant but certainly not spectacular results.  The two main questions
that we address are:

\begin{itemize}

\item Do X-ray sources (and the XRB) trace mass in the Universe and
what is their bias parameter?

\item What are the best observational approaches to obtain information on
the large-scale structure of the Universe at intermediate redshifts with
X-rays?

\end{itemize}

Except when otherwise stated we use $H_0=100\, h\, {\rm km}\, {\rm
s}^{-1}\, {\rm Mpc}^{-1}$, $q_0=0.5$ and $\Lambda=0$.

\section {The X-ray sky on the largest scales}

The distribution of the XRB fluctuations on the largest scales and their
link to inhomogeneities in the distribution of matter 
has been an active field of research for many years. The observational
resources have been mostly limited to the HEAO-1 A2 experiment which
scanned the sky with a resolution of $3^{\circ}\times 1.5^{\circ}$ at
photon energies 2-60 keV.   

\subsection {The dipole of X-ray sources}

Since the Galaxy is moving with respect to the frame
where the CMB would be isotropic towards $l=264^{\circ},
b=48^{\circ}$, there must be an overdensity of sources which are pulling
us towards that direction.  The distribution of X-ray sources in the
sky should therefore exhibit an approximate large-scale dipolar distribution
pointing towards the same direction. 

Using the AGNs in the Piccinotti et al (1982) flux-limited sample of
X-ray sources (2-10 keV flux limit $\sim 3\times 10^{-11}\, {\rm
erg}\, {\rm cm}^{-2}\, {\rm s}^{-1}$), Miyaji \& Boldt (1990) and
Miyaji (1994) found the dipole of these sources to point towards
$l=318^{\circ}, b=38^{\circ}$ with a large error circle ($\sim
30^{\circ}$ radius). The dipole appears to saturate at $50-100\,
h^{-1}\, {\rm Mpc}$ and is roughly aligned with the CMB dipole.
Within the framework of linear theory, this allows the bias parameter
of the X-ray selected AGN to be estimated, giving a somewhat large
value ($b_X\Omega_0^{-0.6}\sim 3-6$). Uncertainties come primarily
from the indetermination of the redshift at which the dipole
saturates.

Plionis \& Kolokotronis (1998) and Kolokotronis et al (1998) have
measured the dipole of an X-ray flux-limited sample of galaxy
clusters.  This is again in rough alignment with the CMB dipole, but
it appears to saturate at $\sim 160 h^{-1}\, {\rm Mpc}$. As
expected in all popular scenarios where clusters arise in extreme
peaks of the underlying dark-matter distribution, they exhibit a large
bias parameter ($b_X\sim 4$, see Table 4).

\begin{table}
\begin{center}
\begin{tabular}{|l|c|c|c|c|l|}
\hline
      & $l\ (deg)$ & $b\ (deg)$ & Err $(deg)$ & XRB ampl (\%) & Ref \\
\hline
CMB & 264 & 48 &    & 0.15 & F96\\
AGN & 318 & 38 & 30 &      & MB90\\
Clusters & 260 & 5  & 15 &  & PK98\\
Soft XRB & 288 & 25 & 19 & 1.7 & PG99\\
Hard XRB & 338 & 47 & 25 & 0.11 & S99\\
\hline
\end{tabular}
\caption {Dipoles of X-ray source populations and the XRB.  References
are Fixsen et al (1996); MB80: Miyaji \& Boldt (1980); PK98: Plionis
\& Kolokotronis (1998); PG99: Plionis \& Georgantopoulos (1999); S99:
Scharf et al (1999).}
\end{center}
\end{table}

The fact that the dipoles of the two most numerous classes of
extragalactic X-ray sources (AGNs and clusters) are roughly aligned
with the CMB dipole is encouraging.  We note, however, that all-sky
deeper samples of these objects (particularly X-ray selected AGN)
would enormously help in defining the distance at which the
contribution to the dipole saturates and therefore in measuring the
bias parameter.

\subsection {The dipole of the X-ray background}

There are two reasons why the XRB should show a dipole signal: our
motion relative to the CMB rest frame (the so-called Compton-Getting
effect) and the excess contribution of the sources that cause this
motion in the same direction.  The XRB dipole is expected to be
aligned with the CMB dipole, but the amplitude should be larger than
the Compton-Getting effect, allowing for the excess emissivity.

There are two basic problems in measuring the XRB dipole: one is the
contribution of the Galaxy and the other one is the integrated nature
of the XRB whereby confusion noise dominates on all angular scales.
Warwick, Pye \& Fabian (1980) realized that even at photon energies
$>2$ keV and galactic latitudes $\mid b \mid > 20^{\circ}$ a residual
galactic contribution $\sim 2-7\%$ is present.  Iwan et al (1982)
modelled this galactic component in terms of a finite radius disk with
thermal spectrum at $T\sim 9$ keV.  To emphasize how difficult is to
obtain the extragalactic signal, the galactic contribution
amounts to a few \% at the galactic poles, while the effect it is
being looked for is less than 1\%.

Attempts to look for singular enhancements of the XRB surface
brightness include those by Warwick et al (1980), the Jahoda \&
Mushotzky (1989) search for emissivity from the great attractor, the
Mushotzky \& Jahoda (1992) search for XRB negative fluctuations
towards the most prominent voids and the unsuccessful detection of
X-ray emission from superclusters by Persic et al (1990). 

By modelling out the Galaxy, Shafer (1983) and Shafer \& Fabian (1983)
found a dipole signal significant at $\sim 2\sigma$ level in the
HEAO-1 A2 map.  Most of the subsequent dipole refinements have used
the same data with increasingly finer corrections for detector drifts
and other unwanted effects.  The latest one is by Scharf et al (1999),
who excluded the galactic plane, the Magellanic clouds and also
regions around the Piccinotti et al (1982) sources, which leaves less
than 50\% of the sky for the dipole analysis.  Various methods are
used to deal with the masked regions (including spherical harmonic
reconstruction) and the results are shown in Table 1.  The dipole
signal is very clearly detected and its intensity appears larger than
the Compton-Getting effect.  The direction of this extra large-scale
structure dipole caused by the fluctuations in the source density is
only roughly aligned with the direction of our motion, and its
amplitude is similar to the Compton-Getting effect as predicted by
theory (Lahav, Piran \& Treyer 1997).

In an analysis of the $ROSAT$ all-sky data (0.9-2.4 keV), Plionis \&
Georgantopoulos (1999) also find a dipole component.  The Galaxy is
modeled according to the Iwan et al (1982) model and they further exclude
other regions associated with the Galaxy.  The direction of the
resulting dipole is in better agreement with the CMB dipole, but the
amplitude is almost a factor of 10 larger than the Compton-Getting
effect.  

There are various reasons for the discrepancy between these
measurements. First, an extra residual contribution from the Galaxy is
likely to contaminate more strongly the $ROSAT$ data than the HEAO-1
A2 data. This would explain why the $ROSAT$ dipole points closer to
the galactic plane and that its amplitude is larger. A second reason
for the discrepancy is the fact that Scharf et al (1999) have excluded
regions around the galaxy clusters present in the Piccinotti et al
(1982) sample (which are known to have a very large bias parameter and
represent 50\% of the extragalactic sources in that sample) but
Plionis \& Georgantopoulos (1999) have not.  In fact these last
authors note that the contribution from the Virgo cluster alone is of
the order of 20\% of the detected dipole.  A good exercise which could
give some insight on the level of the galactic contamination in the
$ROSAT$ data would be to exclude the clusters in the $ROSAT$ analysis
and not excise the Piccinotti et al (1982) sources from the analysis
of the A2 data.

\subsection {Higher order multipoles of the X-ray background}

Lahav, Piran \& Treyer (1997) proposed the use of a multipole
expansion of the angular variations of the XRB in order to measure the
large-scale structure of the Universe. Under fairly general
assumptions, the coefficients $a_{lm}$ of the harmonic expansion would be the
sum of a large-scale structure term $a_{lm}^{(LLS)}\propto l^{-0.4}$ and a
confusion noise term which is a function of the flux $S_{cut}$
down to which sources have been excised from the maps for the
multipole analysis $a_{lm}^{(Noise)}\propto S_{cut}^{\gamma-1}$, where
$\gamma$ is the slope of the {\it integral} source counts in the
energy band used ($N(>S)\propto S^{-\gamma}$). 

Treyer et al (1998) performed this analysis on the HEAO-1 A2 all sky
data by removing regions around the Piccinotti et al (1982) sample and
the galactic plane.  They find evidence for a growth of the spherical
harmonic coefficients growing at low values of $l$ in a manner roughly
consistent with the predictions.  The significance of the signal is
difficult to assess as the harmonic coefficients are not independent
due to cross-talk between different orders introduced by the
masking. Assuming a redshift dependent bias parameter for the X-ray
sources parametrized as $b_X(z)=b_X(0)+z[b_X(0)-1]$ (which assumes
that all galaxies form at some past epoch, Fry 1996), they estimate a
rather modest bias parameter ($1.0<b_X(0)<1.6$).  In their diagrams it
is also seen that the dipole ($l=1$) has an unusually large amplitude
compared to higher harmonics.

The way to go is indeed to have precise measurements of the XRB
intensity on large angular scales, but with the possibility of
excluding sources down to the faintest possible levels.  Treyer et al
(1998) suggest that an all-sky map with XRB intensities measured with
a 1\% precision and with sources excised down to $3\times 10^{-13}\,
{\rm erg}\, {\rm cm}^{-2}\, {\rm s}^{-1}$ (i.e., 100 times fainter
than the Piccinotti et al catalogue) would be ideal for a spherical
harmonic analysis.

\section {Cross-correlations of galaxy catalogues with XRB
intensities}

An alternative way that has been devised to look for structure in the
X-ray sky is to cross-correlate the unresolved XRB intensity with
catalogues of galaxies. The amplitude of the cross-correlation
function (CCF) between the X-ray intensity $I_{XRB}$ and the galaxy surface
density $N_g$, $W_{Xg}(\theta)=\langle I_{XRB}N_g\rangle_{\theta}/\langle
I_{XRB}\rangle\langle N_g\rangle-1$ at zero-lag ($\theta=0$) provides an
approximate measurement of the fraction of the XRB arising either in
the catalogued galaxies or in sources clustered with them within a
scale of the beam with which X-ray observations have been obtained
(Lahav et al 1993).  

\begin{table}
\begin{center}
\begin{tabular}{|l|l|c|l|}
\hline
XRB data & Galaxy catalogue & $W_{XG}(0)$ & Ref \\
\hline
HEAO-1 A2 & IRAS 2Jy & $7\times 10^{-3}$ & M94\\
GINGA NGP & IRAS 0.7Jy & $1.4\times 10^{-2}$ & C95\\
GINGA NGP & UGC & $1.1\times 10^{-2}$ & C95\\
HEAO-1 A2 & IRAS 12$\mu$ all & $<9.6\times 10^{-3}$ & B95\\
HEAO-1 A2 & IRAS 12$\mu$ Sy1 & $1.1\times 10^{-1}$ & B95\\
HEAO-1 A2 & IRAS 12$\mu$ Sy2 & $3.1\times 10^{-2}$ & B95\\
\hline
\end{tabular}
\caption {Cross-correlation signal at zero-lag ($W_{Xg}$) of galaxy
catalogues with XRB data either from the HEAO-1 A2 experiment or the
GINGA North Galactic Pole scans.  References are M94: Miyaji et al 
(1994); C95: Carrera et al 1995; (B95): Barcons et al (1995).}
\end{center}
\end{table}

Positive signals have been found for $W_{Xg}$, typically of the order
of 1\% when the galaxies are optically or infrared selected, and up
to $>10$\% when active galaxies are selected (see table 2). The
interpretation of this signal requires to model the clustering of
X-ray sources around the catalogued galaxies, which is indeed
modulated by the bias parameter $b_X$.  Using $b_X=1$, it is found
that the local volume emissivity of optically selected galaxies
amounts to $\sim 10^{39}\, h\, {\rm erg}\, {\rm s}^{-1}\, {\rm
Mpc}^{-3}$ (Lahav et al 1993; Miyaji et al 1994; Carrera et al 1995),
most of which is contributed by Seyfert galaxies and QSOs (Barcons et
al 1995).

When this volume emissivity is extrapolated to higher redshifts, the
fraction of the XRB intensity due to the precursors of the catalogued
galaxies can be predicted (Lahav
et al 1993).   Carrera et al (1995) find 
that 10-30\% of the hard X-ray background might be produced by
optically selected galaxies without exceeding the upper limits on the
autocorrelation function of the XRB.   This value is similar to the result
of cross-correlation analyses of deep $ROSAT$ X-ray images with deep
optical images in the same fields (Almaini et al 1997).

A constraint on the bias parameter of X-ray sources from the CCF
results can be derived by taking into account that a fraction
$f\approx 2/3$ of the CCF signal arises from sources clustered with
the catalogued galaxies. As that contribution scales linearly with
$b_X$, the local volume emissivity scales $\propto {3\over
1+2b_X}$. Since the AGN-only local emissivity is also $\sim 10^{39}\, {\rm
erg}\, {\rm cm}^{-2}\, {\rm s}^{-1}$, we can safely derive that
$b_X<2$ as otherwise the total volume emissivity will be significantly
less than the AGN emissivity.

\section {Clustering of X-ray selected sources}

In the recent years large, complete samples of X-ray selected AGN have
been built.  This has allowed the direct measurement of the 3D spatial
correlation function $\xi(r)$ for these objects and its comparison
with the spatial correlation function of galaxies selected at other
wavebands.

Carrera et al (1998) used two complete samples of X-ray selected AGN
in pencil beam survey regions, spanning a wide redshift range
($0<z<2$) to search for clustering signals and deriving its amplitude
and redshift evolution.  Clustering is found to be 99\% significant at
$z<1$.  When the spatial correlation function is fitted to a standard
power-law form $\xi(r)=(1+z)^p({r\over r_0})^{-1.8}$ (for comoving
$r$), it is seen that comoving or slower clustering evolution is
excluded, and that even for stable or linear growth the values of
$r_0$ permitted by the data are of the same order as the ones derived
from clustering of $IRAS$ galaxies.  Carrera et al (1998) conclude
that X-ray selected AGN are not significantly biased $0.7<b_X<2$.

Akylas, Plionis \& Georgantopoulos (1999) have used the $ROSAT$ All
Sky Survey sources to derive a local angular correlation function
from which they estimate a somewhat higher correlation length
($r_0\sim 7-9\, {\rm Mpc}$), consistent with optically selected QSO
clustering (La Franca et al 1998) and comoving clustering evolution.
The obvious weakness of this method is that it is not based on 3D but
2D data.  

\section {Fluctuations and Anisotropies in the XRB}

The method of auto-correlating the XRB intensity at various
separations has been extensively used in an effort to detect small
scale structure in the XRB attributable to source clustering (Barcons
\& Fabian 1989, De Zotti et al 1990, Jahoda \& Mushotzky 1991, Carrera
et al 1991, Carrera \& Barcons 1992, Carrera et al 1993, Chen et al
1994). These works produced a set of upper limits for the
auto-correlation function of the XRB $W_{XX}(\theta)=\langle
I_{XRB}I_{XRB}\rangle_{\theta}/\langle I_{XRB}\rangle^2-1$ on
different angular scales (except for the Jahoda \& Mushotzky 1991
work, which claimed a detection at separations $\sim 10^{\circ}$) of
the order of $10^{-3}-10^{-4}$ which constrained the clustering
properties of the underlying source population (see, e.g., Fabian \&
Barcons 1992).

Under the assumption of comoving clustering evolution, the sources of
the XRB cannot be more strongly clustered than optically selected
galaxies (see Carrera \& Barcons 1992), in which case $b_X\sim 1$.
However, as explained above, Carrera et al (1998) found marginal
evidence for faster clustering evolution in samples of X-ray selected
AGN.  This means that $b_X$ could be higher without violating the
upper limits on the autocorrelation function, as the sources that
produce the bulk of the XRB at high redshift could be very weakly
clustered.

A further method employing the XRB angular variations has been to
search for fluctuations in the XRB intensity distribution in excess of
the ones expected from confusion noise produced by unresolved
sources. These excess fluctuations should then be attributed to source
clustering if all remaining noises (counting noise, systematics, etc.)
could be removed. Studies of this kind have invariably lead to upper
limits summarized in Table 3. What actually limits the sensitivity of
this method is the statistics: it scales as $N_{obs}^{-1/2}$ where
$N_{obs}$ is the number of independent measurements of the XRB
intensity that have been used to derive the excess fluctuations.

\begin{table}
\begin{center}
\begin{tabular}{|l|c|c|l|}
\hline
XRB data & Beam & $\left( {\Delta I\over I}\right)_{excess}$ & Ref\\
\hline
HEAO-1 A2 & $5^{\circ}\times 5^{\circ}$& $<0.02$ & S83 \\
Ginga& $2^{\circ}\times 1^{\circ}$& $<0.04$ & B97\\
ROSAT& $\pi \times (2.5')^2$ & $<0.07$&CFB97\\
ROSAT& $\pi (10'^2-5'^2)$ & $<0.12$ &CFB97\\
\hline
\end{tabular}
\caption {Upper limits to excess fluctuations. References are S83:
Shafer (1983); B97: Butcher et al (1997); CFB97: Carrera, Fabian \&
Barcons (1997)}
\end{center}
\end{table}

Excess fluctuations are related to the power spectrum of the density
field of the Universe, weighted with the X-ray volume emissivity as a
function of redshift (Barcons, Fabian \& Carrera 1998).  The method is
potentially very powerful as it reflects the clustering properties of
the sources that produce the bulk of the XRB at redshifts $z>1$.

\section {The bias parameter of X-ray sources}

Over the past sections we have discussed various approaches to detect
and measure the clustering properties of X-ray sources.  Table 4
summarizes the inferred bias parameter $b_X$ from these
studies. Measurements are carried out with a variety of methods,
correspond to different objects, are sensitive to different redshifts
and also to different scales.  Besides that, all dynamical estimates
actually measure the combination $b_X\Omega_0^{-0.6}$.

Measurements of the correlation function are also affected by the
cosmological parameters in the computation of the distances at
significant redshifts, beyond the obvious linear dependence on $H_0$.
If we live in an accelerating Universe, the Carrera et al (1998)
correlation length would have to be scaled up by 30-50\%, resulting in
a subsequent increase of almost a factor of 2 in the bias parameter.
Given the uncertainties in the values of $q_0$ and $\Lambda$ (even for
a flat Universe), the Carrera et al (1998) and Akylas et al (1999) results
cannot be considered inconsistent.

\begin{table}
\begin{center}
\begin{tabular}{|l|l|c|c|}
\hline
Measurement & Reference& Scale (Mpc) &$b$ ($\Omega_0=1$)\\
\hline
X-ray cluster dipole & PK98 & 10-100& 4\\
X-ray AGN dipole & MB90 & 1000& 3-6\\
XRB-galaxy CCF & & 10-100 & $<2$\\
XRB dipole vs bulk motions & S99 & 1000& 2-7\\
XRB multipoles vs bulk motions & T98 & 100-1000& 1-2\\
Clustering of distant AGN & C98 & 10-100& 1-2 \\
Clustering of nearby AGN & APG99 &    100& 2-3 \\
\hline
\end{tabular}
\end{center}
\caption{Bias parameters as inferred from various measurements.
references are: PK98 Plionis \& Kolokotronis (1998); MB90: Miyaji \&
Boldt (1990); S99: Scharf et al
(1999); T98: Treyer et al (1998); C98: Carrera et al (1998); APG99: Akylas,
Plionis \& Georgantopoulos (1999).}
\end{table}

As expected, clusters are a largely biased population ($b_X\sim 4$)
compared to AGN ($b_X\sim 1-2$). The multipoles of the XRB are
expected to be dominated by AGN, as these objects are the main sources
of the XRB.  The bias parameter derived from the XRB multipoles is
consistently in agreement with the bias parameter derived from AGN
clustering ($b_X\sim 1-2$).  The exception to this is the XRB dipole
which implies a larger value of $b_X$.  This could be partly due to a
larger cluster contribution, as the lowest order multipoles are most
sensitive to nearest (and brightest) sources, where the cluster
contribution to the source counts ($\sim 10\%$ on average in the deep
extragalactic surveys) is $\sim 50\%$ for the Piccinotti et al (1982)
sample.

\section {Future Prospects}

X-ray astronomy is now in a position to address cosmological studies.
X-ray selected AGN which produce most of the X-rays in the Universe,
appear to trace mass with a moderate bias parameter $b_X\sim 1-2$, but
that has to be better defined as a function of scale and redshift.
$Chandra$ and XMM will carry out several deep `pencil beam' surveys
which, after subsequent identification of the serendipitous sources
discovered, will define the redshift evolution of the AGN X-ray
luminosity function at photon energies $>2 {\rm keV}$ and therefore the
X-ray volume emissivity as a function of redshift. However, these
surveys will not map sufficiently large areas of the sky which are
necessary to trace the large-scale structure of the Universe at the
redshifts where the XRB was produced.

The obvious way to go would be to survey very large areas of the sky
(the whole sky even better) for X-ray sources, in order to have a most
complete picture.  Unless hard X-rays are produced at significantly
lower redshifts than soft X-rays (which is doubtful in view of the
$ASCA$ and {\it BeppoSAX} surveys), to reach $z\sim 1$ where a
significant fraction of the X-ray emissivity in the Universe resides,
these surveys will have to go at least down to $\sim 10^{-14}\, {\rm
erg}\, {\rm cm}^{-2}\, {\rm s}^{-1}$.

There is an alternative which is to perform high sensitivity
observations of the XRB with a beam corresponding to the linear scale
to be probed (Barcons, Fabian \& Carrera 1998).  As the peak of the
power spectrum of the density field of the Universe occurs at comoving
wavenumbers $\sim 0.01-0.1\, h \, {\rm Mpc}^{-1}$, for a standard
geometry a $1^{\circ}$ resolution is well matched to this at $z\sim
1-3$. All-sky measurements of the XRB intensity on that angular scale with a
precision of a few \% could then be used to detect the excess
fluctuations due to source clustering which are expected to be just
below 1\% in amplitude.  Controlling all other
possible sources of excess fluctuations well below that level requires
a stable large-area detector (to reduce photon counting noise) and
probably an X-ray monitor which images simultaneously the brightest
sources in the field.


\begin{acknowledgements}
Partial financial support for this work was provided by the DGESIC
under project PB95-0122.
\end{acknowledgements}

\end{document}